\input psfig.sty
\nopagenumbers 
\magnification=\magstep1 
\hsize 6.0 true in 
\hoffset 0.25 true in 
\emergencystretch=0.6 in 
\vfuzz 0.4 in 
\hfuzz 0.4 in 
\vglue 0.1true in 
\mathsurround=2pt 
\def\nl{\noindent} 
\def\nll{\hfil\break\noindent} 
\def\np{\hfil\vfil\break} 
\def\ppl#1{{\leftskip=9cm\noindent #1\smallskip}} 
\def\title#1{\bigskip\noindent\bf #1 ~ \trr\smallskip} 
 
\font\trr=cmr10 
\font\bf=cmbx10 
\font\bmf=cmmib10 
\font\bmfs=cmmib8 
\font\bmfg=cmmib10 scaled 1500 
\font\sl=cmsl10 
\font\it=cmti10 
\font\trbig=cmbx10 scaled 1500 
\font\tiny=cmr8 
\def\mb#1{\hbox{\bmf#1}} 
\def\mbs#1{\hbox{\bmfs#1}} 
\def\ng{>\kern -9pt|\kern 9pt} 
\def\hi#1#2{$#1$\kern -2pt-#2} 
\def\hy#1#2{#1-\kern -2pt$#2$} 

\def\half{{1 \over 2}} 
 
\def\rtitle{RELATIVISTIC \hi{N}{BOSON} SYSTEMS} 
\def\ptitle{RELATIVISTIC {\bmfg N}-BOSON SYSTEMS} 
\def\ptitlee{BOUND BY OSCILLATOR PAIR POTENTIALS} 
\def\dbox#1{\hbox{\vrule 
\vbox{\hrule \vskip #1\hbox{\hskip #1\vbox{\hsize=#1}\hskip #1}\vskip #1 
\hrule}\vrule}} 
 
\def\qed{\hfill \dbox{0.05true in}} 
\output={\shipout\vbox{\makeheadline\ifnum\the\pageno>1 {\hrule} \fi 
{\pagebody}\makefootline}\advancepageno} 
 
\headline{\noindent {\ifnum\the\pageno>1 
{\tiny \rtitle\hfil page~\the\pageno}\fi}} 
\footline{} 
\newcount\zz \zz=0 
\newcount\q 
\newcount\qq \qq=0 
 
\def\pref #1#2#3#4#5{\frenchspacing \global \advance \q by 1 
\edef#1{\the\q}{\ifnum \zz=1 { %
\item{[\the\q]}{#2} {\bf #3},{ #4.}{~#5}\medskip} \fi}} 
 
\def\bref #1#2#3#4#5{\frenchspacing \global \advance \q by 1 
\edef#1{\the\q}{\ifnum \zz=1 { %
\item{[\the\q]}{#2}, {\it #3} {(#4).}{~#5}\medskip} \fi}} 
 
\def\gref #1#2{\frenchspacing \global \advance \q by 1 
\edef#1{\the\q}{\ifnum \zz=1 { %
\item{[\the\q]}{#2}\medskip} \fi}}

\def\sref #1{~[#1]} 
\def\srefs#1#2{~[#1--#2]} 
 
\def\references#1{\zz=#1 
\parskip=2pt plus 1pt 
{\ifnum \zz=1 {\noindent \bf References \medskip} \fi} \q=\qq 
\pref{\bse}{E.~E.~Salpeter and H.~A.~Bethe, Phys.~Rev.}{84}{1232 (1951)}{} 
\pref{\se}{E.~E.~Salpeter, Phys.~Rev.}{87}{328 (1952)}{} 
\bref{\lieb}{E.~H.~Lieb and M.~Loss}{Analysis}{American Mathematical Society, 
New York, 1996} {The definition of the Salpeter kinetic-energy operator is 
given on p.~168.} 
\pref{\luca}{W.~Lucha and F.~F.~Sch\"oberl, Phys.~Rev.\ A}{54}{3790 (1996)}{} 
\pref{\hallad}{R. L. Hall, W.~Lucha, and F.~F.~Sch\"oberl, J.~Phys.\ 
A}{34}{5059 (2001)}{} 
\pref{\hous}{W. M. Houston, Phys.~Rev.}{47}{942 (1935)}{} 
\pref{\post}{H. R. Post, Proc. Phys. Soc. London}{66}{942 (1953)}{}
\pref{\hallaa}{R.~L.~Hall, Phys.~Rev.\ A}{51}{3499 (1995)}{}
\pref{\lucq}{W.~Lucha and F.~F.~Sch\"oberl, Phys.~Rev.\ A}{60}{5091 (1999)}{} 
\pref{\lucacc}{W.~Lucha and F.~F.~Sch\"oberl, Int.~J.~Mod.~Phys.\ A}{15}{3221 
(2000)}{} 
\pref{\hallab}{R.~L.~Hall, Can.~J.~Phys.}{50}{305 (1972)}{} 
\pref{\hallac}{R.~L.~Hall, Aequ.~Math.}{8}{281 (1972)}{} 
\pref{\halla}{R.~L.~Hall, J.~Math.~Phys.}{24}{324 (1983)}{}
\pref{\hallb}{R.~L.~Hall, J.~Math.~Phys.}{25}{2708 (1984)}{}
\pref{\hallae}{R.~L.~Hall, W.~Lucha, and F.~F.~Sch\"oberl, J.~Math.~Phys.}{42}{5228 (2001)}{} 
\pref{\lucb}{W.~Lucha and F.~F.~Sch\"oberl, Int.~J.~Mod.~Phys.\ A}{14}{2309 
(1999)}{} 
\bref{\fel}{W. Feller}{An introduction to probability theory and its 
applications, Volume II}{John Wiley, New York, 1971} {Jensen's inequality is 
discussed on p.~153.} 
 
} 
 
\references{0} 
 
\topskip=20pt 
\trr 
\ppl{CUQM-87}\ppl{HEPHY-PUB 744/01}\ppl{UWThPh-2001-34}\ppl{math-ph/0110015} 
\ppl{May 2003}\medskip 
\vskip 0.4 true in 
\centerline{\trbig \ptitle} 
\medskip 
\centerline{\trbig \ptitlee} 
\vskip 0.4true in 
\baselineskip 12 true pt 
\centerline{\bf Richard L.~Hall$^1$, Wolfgang Lucha$^2$, and Franz 
F.~Sch\"oberl$^3$}\medskip 
\nll $^{(1)}${\sl Department of Mathematics and Statistics, Concordia 
University, 1455 de Maisonneuve Boulevard West, Montr\'eal, Qu\'ebec, Canada 
H3G 1M8} 
\nll $^{(2)}${\sl Institut f\"ur Hochenergiephysik, \"Osterreichische 
Akademie der Wissenschaften, Nikolsdorfergasse 18, A-1050 Wien, Austria} 
\nll $^{(3)}${\sl Institut f\"ur Theoretische Physik, Universit\"at Wien, 
Boltzmanngasse 5, A-1090 Wien, Austria} 
  
\nll{\sl rhall@mathstat.concordia.ca, wolfgang.lucha@oeaw.ac.at, 
franz.schoeberl@univie.ac.at} 
\bigskip\medskip 
\baselineskip = 18true pt 
 
\centerline{\bf Abstract}\medskip 
 
\noindent We study the lowest energy $E$ of a relativistic system of $N$ 
identical bosons bound by harmonic-oscillator pair potentials in three 
spatial dimensions. In natural units $\hbar=c=1$ the system has the 
semirelativistic ``spinless-Salpeter'' Hamiltonian$$H= 
\sum_{i=1}^N\sqrt{m^2+\mb{p}_i^2}+\sum_{j>i=1}^N\gamma|\mb{r}_i-\mb{r}_j|^2, 
\quad\gamma>0.$$We derive the following energy bounds: 
$$E(N)=\min_{r>0}\left[N\left(m^2+{{2(N-1)P^2}\over{Nr^2}}\right)^\half 
+{N\over 2}(N-1)\gamma r^2\right],\quad N\ge 2,$$where $P=1.376$ yields a 
lower bound and $P=3/2$ yields an upper bound for all $N\ge 2.$ A sharper 
lower bound is given by the function $P(\mu),$ where $\mu = m(N/(\gamma(N-1)^2))^{1\over 3},$
 which makes the formula for 
$E(2)$ exact: with this choice of $P,$ the bounds coincide for all $N\ge 2$ 
in the Schr\"odinger limit $m\to\infty$. 
\np 
 
\title{I.~~Introduction and Main Result} 
 
Many-body problems form essential links between quantum-theoretical models 
and real nuclear, atomic, or macroscopic systems. However, even for 
nonrelativistic quantum theory, there are very few many-body problems that 
have explicit analytic solutions; the harmonic oscillator and the attractive 
delta interaction are well-known exceptions. In relativistic quantum theories 
the situation is even worse, in spite of~the fact that the phenomenon of 
particle creation allowed by quantum field theory would suggest that there is 
no such thing as a one-body problem in that theory. Therefore, it is of 
considerable interest to study model $N${\kern-1pt}-body systems within the 
framework of the semirelativistic ``spinless-Salpeter'' equation. For this 
problem there exists a well-defined nonrelativistic limit which yields a 
useful consistency check. Specifically, we investigate in this paper the 
relative energy $E$ of a system of $N$ identical bosons represented by a 
semirelativistic ``spinless-Salpeter'' Hamiltonian\sref{\bse,\se} of the 
form$$H=\sum_{i=1}^N\sqrt{m^2+\mb{p}_i^2} 
+\sum_{j>i=1}^N\gamma|\mb{r}_i-\mb{r}_j|^2,\eqno{(1.1)}$$where $m$ is the 
boson mass, and $\gamma>0$ is a coupling parameter, and we have chosen units 
in which $\hbar=c=1.$ The operators $\mb{p}_i$ are defined\sref{\lieb,\luca} 
in the momentum-space representation where they become multiplicative 
operators (\kern -2pt\hi{c}{variables}). The present work is an extension to 
the case of $N$ bosons of our earlier study\sref{\hallad} in which we derived 
energy bounds for the corresponding \hi{1}{body} problem. We may compare $H$ 
with the corresponding Schr\"odinger \hi{N}{body} problem with Hamiltonian 
$$H_{\rm S}=\sum_{i=1}^N{{\mb{p}_i^2}\over{2m}} 
+\sum_{j>i=1}^N\gamma|\mb{r}_i-\mb{r}_j|^2.\eqno{(1.2)}$$
\nl Given our goal of 
investigating the {\it relative\/} (that is, binding) energies, both of these 
Hamiltonians have the unwelcome feature that they include the kinetic energy 
of the center-of-mass motion. This is easy to remedy for $H_{\rm S},$ but a 
correct form is not so immediate in the relativistic case $H.$ The exact 
solution to the \hi{N}{body} harmonic-oscillator problem is periodically 
``rediscovered'' but has been known at 
least since 1935 when Houston\sref{\hous} solved it.  Later, Post\sref{\post} 
studied the non-relativistic translation-invariant problem: the 
exact ground-state energy $E_{\rm S}$ may be expressed\sref{\hallaa}
 for $N\ge 2$ (in three dimensions) by the simple formula
$$\varepsilon=3v^\half,\quad\varepsilon={{mE_{\rm S}}\over{N-1}},\quad 
v={{mN\gamma}\over 2}.\eqno{(1.3)}$$
\nl Thus $\varepsilon$ is exactly the bottom 
of the spectrum of the \hi{1}{body} Hamiltonian $-\Delta+vr^2.$ In this paper 
we shall prove the following statement.\smallskip\nll{\bf Theorem~1}\nll{\it 
Bounds on the ground-state energy eigenvalue $E$ of the semirelativistic 
Hamiltonian (1.1) are provided by the formula 
$$E=\min_{r>0}\left[N\left(m^2+{{2(N-1)P^2}\over{Nr^2}}\right)^\half+{N\over 
2}(N-1)\gamma r^2\right],\quad N\ge 2,\eqno{(1.4)}$$
\nl which yields an upper 
bound on $E$ when $P=3/2,$ and a lower bound on $E$ when $P=P(\mu),$
where $\mu = m(N/(\gamma(N-1)^2))^{1\over 3},$ a function 
that makes the approximation (1.4) exact in the case $N=2.$ The function 
$P(m)$ is monotone increasing with $m,$ has bounds$$1.376<P(m)<{3\over 
2},\eqno{(1.5)}$$and has the limit
$$\lim_{m\to\infty}P(m)={3\over 
2}.\eqno{(1.6)}$$
In the \hy{large}{m} limit, the upper and lower bounds 
coalesce to the corresponding exact (nonrelativistic) Schr\"odinger energy 
$E_{\rm NR}=E_{\rm S}+Nm.$}\smallskip 
 
The paper is primarily concerned with proving Theorem~1. The main technical 
difficulties are twofold: to keep the fundamental symmetries of translation 
invariance and boson permutation symmetry, and to find ways of 
``penetrating'' the square-root operator of the Salpeter kinetic energy. Our 
policy is to work with Jacobi relative coordinates to guarantee translation 
invariance of the wave functions, and to accept the concomitant complications 
of permutation symmetry. We discuss the relative coordinates and some of 
their properties in Sec.~II. We shall exploit the necessary permutation 
symmetry to relate the \hi{N}{body} energy to that of a scaled and reduced 
\hi{2}{body} problem. The exact solution of the \hi{1}{body} problem is 
discussed in Sec.~III. It is well known that the \hi{1}{body} Salpeter 
problem is equivalent to a Schr\"odinger problem with Hamiltonian 
$-\Delta+\sqrt{m^2+r^2}$\sref{\lucq,\lucacc}. We take the position in this 
paper that the lowest eigenvalue $e(m)$ of this problem, which is easy to 
find numerically, is at our disposal. In Fig.~1 we exhibit graphs of the 
functions $\{e(m),P(m)\}.$ The extension of these results to the \hi{2}{body} 
problem is treated in Sec.~IV. The lower bound discussed in Sec.~V is 
rendered possible by an operator property introduced~in Sec.~II that allows 
us, in a sense, to remove certain annihilation operators from~inside the 
square-root operator. For the \hi{N}{body} upper bound discussed in Sec.~VI 
we~use~a Gaussian wave function and minimize the energy expectation with 
respect to a scale variable. The calculation is helped by special factoring 
properties of the Gaussian and by the use of Jensen's inequality. The bounds 
corresponding to $P=\{1.376,1.5\}$ are depicted in Fig.~2, and the 
convergence of the bounds $P=\{P(\mu),3/2\}$ with increasing $m$ 
(where $\mu = m(N/(\gamma(N-1)^2))^{1\over 3}$) 
is shown in Fig.~3, for $2\le N\le 8.$ 
 
\title{II.~~Relative Coordinates} 
 
Jacobi relative coordinates may be defined with the aid of an orthogonal 
matrix $B$ relating the column vectors of the new $[\rho_i]$ and old 
$[\mb{r}_i]$ coordinates according~to$$[\rho_i]=B[\mb{r}_i].\eqno{(2.1)}$$The 
first row of $B$ defines a center-of-mass variable with every entry 
$1/\sqrt{N},$ the second row defines a pair distance $\rho_2 
=(\mb{r}_1-\mb{r}_2)/\sqrt{2},$ and the $k\!$th row, $k\ge 2,$ has the first 
$k-1$ entries $B_{ki}=1/\sqrt{k(k-1)},$ the $k\!$th entry 
$B_{kk}=-\sqrt{(k-1)/k},$ and the remaining entries zero. We define the 
corresponding momentum variables~as$$[\pi_i]=(B^{-1})^{\rm 
t}[\mb{p}_i]=B[\mb{p}_i].\eqno{(2.2)}$$These coordinates have some nice 
properties which we shall need. Firstly, we have 
$$k\sum_{i=2}^k\rho_i^2=\sum_{j>i=1}^k(\mb{r}_i-\mb{r}_j)^2,\quad 
k=2,3,\dots,N,\eqno{(2.3)}$$and similarly for the momenta 
$$k\sum_{i=2}^k\pi_i^2=\sum_{j>i=1}^k(\mb{p}_i-\mb{p}_j)^2,\quad 
k=2,3,\dots,N.\eqno{(2.4)}$$ 
 
It follows immediately that if $\Psi$ is a translation-invariant wave 
function which is symmetric (or antisymmetric) under the permutation of the 
individual-particle indices, then it follows 
that$$\left(\Psi,\rho_i^2\Psi\right)=\left(\Psi,\rho_2^2\Psi\right),\quad 
2\le i\le N,\eqno{(2.5)}$$and 
$$\left(\Psi,\pi_i^2\Psi\right)=\left(\Psi,\pi_2^2\Psi\right),\quad 2\le i\le 
N.\eqno{(2.6)}$$These expectation symmetries might suggest that the wave 
function $\Psi$ is symmetric under permutation of the relative coordinates; 
but this stronger property is not generally true; 
it is the case for Gaussian wave functions. 
Moreover, Gaussian boson wave functions of Jacobi relative coordinates 
uniquely\sref{\hallab, \hallac} have the further factoring property that 
$$\Phi(\rho_2,\rho_3,\dots,\rho_N) 
=\phi(\rho_2)\theta(\rho_3,\dots,\rho_N),\eqno{(2.7)}$$where $\phi$ and 
$\theta$ are also Gaussian. 
 
\title{III.~~The 1-Body Problem} 
 
We consider the \hi{1}{body} problem with Hamiltonian 
$$H_1=\sqrt{m^2+\mb{p}^2}+r^2\quad\to\quad e(m),\eqno{(3.1)}$$where, for 
coupling $\gamma=1,$ $e(m)$ is the lowest eigenvalue as a function of the 
mass $m.$ By transforming this problem into momentum space we obtain the 
equivalent problem$$\tilde H_1=-\Delta+\sqrt{m^2+r^2}\quad\to\quad 
e(m).\eqno{(3.2)}$$Since this Schr\"odinger problem is easy to solve 
numerically to arbitrary accuracy, we shall take the position that $e(m)$ is 
``known'' and at our disposal. We note that in the \hy{large}{m} 
(nonrelativistic or Schr\"odinger) limit, we have$$e(m)\simeq e_{\rm NR}(m)= 
m+{{3}\over{(2m)^\half}}.\eqno{(3.3)}$$We now define, for a given value of 
$m,$ the (lowest) ``kinetic potential''\srefs{\halla}{\hallae}~$\bar{h}(s)$ 
associated with the relativistic-kinetic-energy square-root operator 
$\sqrt{m^2+\mb{p}^2}$ and the harmonic-oscillator potential $r^2$ 
by$$\bar{h}(s)=\inf_{{{\psi\in{\cal 
D}(H_1)}\atop{\|\psi\|=1}}\atop{\left(\psi,\sqrt{m^2+\mbs{p}^2}\psi\right)=s}} 
\left(\psi,r^2\psi\right),\eqno{(3.4)}$$where $\psi(\mb{r})$ is a wave 
function in the domain ${\cal D}(H_1)$ of $H_1.$ That is to say, we find the 
minimum mean-value of the potential, subject to the constraint that the mean 
kinetic energy is held constant at the value $s.$ It follows that the 
eigenvalue may now be recovered from $\bar{h}(s)$ by a further minimization 
with respect to the kinetic energy $s.$ Thus we have 
$$e(m)=\min_{s>m}\left[s+\bar{h}(s)\right].\eqno{(3.5)}$$
\nl It may be difficult to find the kinetic potential $\bar{h}(s)$
 exactly from (3.4).  Instead we construct an effective kinetic potential
$\bar{h}_{\rm eff}(s)$ which, when substituted in (3.5), yields $e(m)$ exactly.  We do this 
by changing the minimization variable from $s>m$ to $r>0$ according to the 
following equations:
$$\bar{h}_{\rm eff}(s)=r^2,\quad 
s=\sqrt{m^2+\left({{P(m)}\over r}\right)^2}.\eqno{(3.6)}$$
Now, by rewriting 
(3.5) in terms of the minimization variable $r$ we obtain the defining relation for $P(m)$ as follows:
$$e(m)=\min_{r>0} 
\left[\sqrt{m^2+\left({{P(m)}\over r}\right)^2}+r^2\right].\eqno{(3.7)}$$
\nl In fact, by inverting (3.7), we find the following expression for $P(m)$ in terms of 
the \hi{1}{body} energy $e(m)$: 
$$P(m)=\left({{2\left(e(m)+\sqrt{e^2(m)+3m^2}\right)}\over{27}}\right)^\half 
\left(2e(m)-\sqrt{e^2(m)+3m^2}\right).\eqno{(3.8)}$$ 
 
The graphs of $e(m)-m$ and $P(m)$ are shown in Fig.~1: both $e(m)$ and $P(m)$ 
are monotone increasing with $m;$ $e(m)-m,$ however, is monotone {\it 
decreasing\/}, in agreement, for large $m,$ with the Feynman--Hellmann 
theorem for the corresponding nonrelativistic case. In the 
(ultrarelativistic) limit $m\to 0$ we have $\tilde H_1\to -\Delta+r,$ that is 
to say, the operator limit is the Schr\"odinger operator for the linear 
potential in three dimensions, with lowest energy $e(0)=2.33810741.$ In the 
(nonrelativistic) \hy{large}{m} limit~we have $H_1\to m-(1/2m)\Delta+r^2,$ 
that is to say, the Schr\"odinger harmonic oscillator with energy $e(m)\simeq 
m + 3/\sqrt{2m}.$ By substituting these ``outer'' energies in (3.8), we 
obtain the bounds
$$1.376<P(m)<{3\over 2}.\eqno{(3.9)}$$ 
 
It is clear from Eq.~(3.7) that the expression for $e(m),$ as a function of 
$m$ {\it and\/} $P,$ is monotone increasing in $P.$ Thus, by substituting, 
respectively, the constants $P=1.376$ and $P=1.5,$ we obtain from this 
formula lower and upper bounds on the \hi{1}{body} energy $e(m).$ These 
bounds agree exactly with the bounds we obtained 
earlier\sref{\hallad,\hallae} for this \hi{1}{body} harmonic-oscillator 
problem.

For later application to the \hi{N}{body} problem, we now consider a more general \hi{1}{body} problem with Hamiltonian
$$H = \beta\sqrt{m^2+\lambda\mb{p}^2}+\gamma r^2\eqno{(3.10)}$$
\nl and positive parameters $\{\beta,\gamma,\lambda\}.$
We find by elementary scaling arguments
 that the eigenvalue $\varepsilon(m,\beta,\gamma\lambda)$ 
corresponding to the operator $H$ may be expressed 
in terms of the energy function $e(m)$ by the explicit formula
$$\varepsilon(m,\beta,\gamma\lambda)= 
\left({{\beta^2}\gamma\lambda}\right)^{1\over 3} 
e\left(m\left({\beta\over{\gamma\lambda}}\right)^{1\over 
3}\right).\eqno{(3.11)}$$ 
\nl In terms of $P,$ we therefore have
$$\varepsilon(m,\beta,\gamma\lambda)=\min_{r>0} 
\left[\beta\left(m^2+\lambda\left({{P}\over r}\right)^2\right)^\half+\gamma 
r^2\right].\eqno{(3.12)}$$
For each $\beta>0,\ \gamma>0,\ \lambda>0,$ this 
formula is therefore exact when 
$$P = P(\mu),\quad {\rm where}\quad \mu = m\left({{\beta}\over{\gamma\lambda}}\right)^{1\over 3},\eqno{(3.13)}$$
\nl it yields a lower bound when $P=1.376,$ and an 
upper bound when $P=1.5.$ As we shall see in the next section,
 the \hi{2}{body} energy is obtained from 
(3.11) or (3.12) by simply setting $\lambda=1,\ \beta=2.$ It is an extension 
of this reasoning that will allow us, in Sec.~V, to obtain also the 
\hi{N}{body}, $N\ge 2,$ lower energy bound by using suitable values for 
$\beta,$ $\gamma,$ and $\lambda.$ 
 
\title{IV.~~The 2-Body Problem} 
 
For the case $N=2$ we have explicitly 
$$H=\sqrt{m^2+\mb{p}_1^2}+\sqrt{m^2+\mb{p}_2^2}+\gamma|\mb{r}_1-\mb{r}_2|^2. 
\eqno{(4.1)}$$Let $\psi(\rho_2)$ be a normalized boson wave function. Then 
the lowest relative eigenvalue of the operator $H$ is the infimum of 
expectation values of the form $(\psi,H\psi).$ But the boson symmetry of 
$\psi(\rho_2)$ means that the two kinetic-energy terms in $(\psi,H\psi)$ must 
have the same value. Moreover, in terms of relative coordinates, the operator 
$\mb{p}_2^2$ may be written$$\mb{p}_2^2={(\pi_1-\pi_2)^2\over 
2}.\eqno{(4.2)}$$Now, the operator $\pi_1$ would immediately annihilate 
$\psi(\rho_2)$ if it were not contained in the square root. We claim that, 
inside the expectation value, the operator $\pi_1$~may simply be removed; 
this may be seen as an immediate generalization of the following observation. 
\smallskip\nll {\bf Lemma~1}\nll{\it Suppose $\Psi(x,y)=\psi(x),$ then 
$$\left[1-\left({{\partial}\over{\partial x}}- {{\partial}\over{\partial 
y}}\right)^2\right]^\half\Psi=\left(1-{{\partial^2}\over{\partial 
x^2}}\right)^\half\Psi.\eqno{(4.3)}$$}\smallskip\nll{\bf Proof of 
Lemma~1}\nll If ${\cal F}$ indicates the 2-dimensional Fourier transform and 
our new variables~are~$\{p, q\},$ then we find ${\cal F}(\Psi)(p,q) 
=\tilde\psi(p)\delta(q),$ and, by definition, the Fourier transform of the 
left-hand side of (4.3) 
becomes$$\left(1+(p-q)^2\right)^\half\tilde\psi(p)\delta(q)= 
\left(1+p^2\right)^\half\tilde\psi(p)\delta(q).\eqno{(4.4)}$$By transforming 
back to the variables $\{x, y\},$ we obtain the right-hand side of (4.3). 
${}$\qed\smallskip 
 
Applying the generalization of this lemma to our problem in three dimensions, 
we find, for $\psi=\psi(\rho_2),$$$(\psi,H\psi)=\left(\psi, 
\left(2\sqrt{m^2+\half\pi_2^2}+2\gamma\rho_2^2\right)\psi\right).\eqno{(4.5)}$$ 
By defining the pair-distance variable 
$\mb{r}=\mb{r}_1-\mb{r}_2=\sqrt{2}\rho_2,$ and the corresponding momentum as 
$\mb{p}=-{\rm i}\nabla_{\mbs{r}}=\pi_2/\sqrt{2},$ we may rewrite (4.5) as 
$$(\psi,H\psi)=\left(\psi, 
\left(2\sqrt{m^2+\mb{p}^2}+\gamma r^2\right)\psi\right).\eqno{(4.6)}$$By 
using a formal relative coordinate $\mb{r},$ we have thus recovered the 
well-known\sref{\lucb} \hi{2}{body} result: the minimum of the right-hand 
side of (4.6) is the bottom of the spectrum of $H$ which corresponds 
precisely to the energy of a \hi{1}{body} problem~with the kinetic-energy 
parameter $\beta=2.$ This result may also be expressed in terms of the 
\hi{1}{body} energy function $e(m)$ by means of Eq.~(3.11). Thus we have 
explicitly~for $N=2$$$E=\left(4\gamma\right)^{1\over 
3}e\left(m\left({2\over\gamma}\right)^{1\over 3}\right).\eqno{(4.7)}$$ 
 
In the next section we shall apply a similar reasoning to the \hi{N}{body} 
problem; however, for $N>2$ we obtain, instead of the exact energy, a lower 
energy bound. 
 
\title{V.~~The Lower Bound} 
 
Suppose that $\Psi(\rho_2,\rho_3,\dots,\rho_N)$ is a normalized 
translation-invariant \hi{N}{boson} wave function. Boson symmetry and, in 
particular, formula (2.3) allow us to write 
$$E\le(\Psi,H\Psi)=N\left(\Psi,\left(m^2+\mb{p}_N^2\right)^\half\Psi\right)+ 
{{N}\choose{2}}\gamma\left(\Psi,2\rho_N^2\Psi\right).\eqno{(5.1)}$$Now, from 
the definition of the relative coordinates, we have$$\mb{p}_N= 
{{1}\over{\sqrt{N}}}\pi_1-\sqrt{{{N-1}\over N}}\pi_N.\eqno{(5.2)}$$ 
Consequently, an application of an immediate generalization of Lemma~1 
allows~us~to ``remove'' the operator $\pi_1$ from the square root of the 
kinetic-energy term and write$$E\le 
N\left(\Psi,\left(m^2+{{N-1}\over{N}}\pi_N^2\right)^\half\Psi\right)+ 
{{N}\choose{2}}\gamma\left(\Psi,2\rho_N^2\Psi\right).\eqno{(5.3)}$$Adapting 
the argument presented in Sec.~IV for the \hi{2}{body} case $N=2,$ we define 
a relative coordinate $\mb{r}=\sqrt{2}\rho_N,$ and the corresponding momentum 
$\mb{p}=\pi_N/\sqrt{2}.$ The expression for the upper bound to the lowest 
\hi{N}{boson} energy $E$ then becomes$$E\le 
N\left(\Psi,\left(m^2+{{2(N-1)}\over{N}}\mb{p}^2\right)^\half\Psi\right)+ 
{{N}\choose{2}}\gamma\left(\Psi,r^2\Psi\right).\eqno{(5.4)}$$The inequality 
(rather than an equality) in (5.4) comes only from the choice of wave 
function. If we find the infimum of such expressions over all normalized 
translation-invariant \hi{N}{boson} wave functions, we would obtain the exact 
energy $E;$ if we find this minimum but without the constraint of boson 
symmetry, then the right-hand side of (5.4) will in general fall below $E$ 
but will in any case be bounded from below by the bottom of the spectrum of 
the \hi{1}{body} semirelativistic Salpeter Hamiltonian 
$$H=N\left(m^2+{{2(N-1)}\over{N}}\mb{p}^2\right)^\half+{{N}\choose{2}}\gamma 
r^2.\eqno{(5.5)}$$
\nl But this latter problem corresponds precisely to Eq.~(3.7) 
if we make the parameter substitutions$$\beta=N,\quad 
\lambda={{2(N-1)}\over{N}},\quad\gamma\to{{N}\choose 
2}\gamma={{N(N-1)}\over{2}}\gamma.\eqno{(5.6)}$$Thus, in view of the $P$ 
representation (3.12), it is clear by choosing $P= P(\mu),$ where 
$\mu = m(N/(\gamma(N-1)^2))^{1\over 3} > 1.376,$ that we have established the lower bound 
(1.4) of Theorem~1. 
 
It is interesting to note that we can also substitute the \hi{N}{body} values 
(5.6) for the parameters $\beta,$ $\gamma,$ and $\lambda$ into the result 
(3.11) for the \hi{1}{body} ground-state energy 
$\varepsilon(m,\beta,\gamma\lambda)$ in order to obtain the following 
explicit expression for the lower 
bound:
$$E\ge\left(N^2(N-1)^2\gamma\right)^{1\over 3} 
e\left(m\left({{N}\over{(N-1)^2\gamma}}\right)^{1\over 
3}\right).\eqno{(5.7)}$$This expression---which is equivalent to the lower 
bound (1.4) of Theorem~1---gives the exact energy and agrees with Eq.~(4.7) 
when $N=2.$ Meanwhile, for all $N\ge 2,$ in the nonrelativistic \hy{large}{m} 
(Schr\"odinger) limit it yields the exact \hi{N}{body} energy$$E_{\rm 
NR}=Nm+3\left({{\gamma}\over{2m}}\right)^\half N^\half(N-1),\eqno{(5.8)}$$ 
reproducing thus the old result of Houston and Post recalled in Eq.~(1.3). 
 
\title{VI.~~The Upper Bound} 
 
For the upper bound we employ a Gaussian wave function of the form 
$$\Phi(\rho_2,\rho_3,\dots,\rho_N)= 
C\exp\left(-\alpha\sum_{i=2}^N\rho_i^2\right),\quad\alpha>0,\eqno{(6.1)}$$where 
$C$ is a normalization constant. The factoring property (2.7) of this 
function and the boson-symmetry reduction leading to (5.4) allows us to write 
$$E\le 
N\left(\phi,\left(m^2+{{2(N-1)}\over{N}}\mb{p}^2\right)^\half\phi\right) 
+{{N}\choose{2}}\gamma\left(\phi,r^2\phi\right),\eqno{(6.2)}$$where the 
function $\phi(r)$ is given 
by$$\phi(r)=\left({{\alpha}\over{\pi}}\right)^{3\over 4}\exp\left(-{{\alpha 
r^2}\over 2}\right).\eqno{(6.3)}$$ 
 
Since the kinetic-energy operator is a {\it concave}\/ function of the square 
$\mb{p}^2$ of the momentum, we can use Jensen's inequality\sref{\fel} to move 
the expectation value $\langle\mb{p}^2\rangle$ inside the square root and 
thus estimate the mean value of this operator from above and write$$E\le 
N\left(m^2+{{2(N-1)}\over{N}}\left(\phi,\mb{p}^2\phi\right)\right)^\half+ 
{{N}\choose{2}}\gamma\left(\phi,r^2\phi\right).\eqno{(6.4)}$$We shall 
minimize this upper bound with respect to the scale variable $\alpha>0.$ We 
parametrize the basic kinetic-energy and potential-energy expectation values 
in terms of a variable $r>0$ by the following relations: 
$$\left(\phi,r^2\phi\right)={{3}\over{2\alpha}}:=r^2,\quad 
\left(\phi,\mb{p}^2\phi\right)={3\alpha\over{2}}=\left({P\over 
r}\right)^2,\quad P:={3\over 2}.\eqno{(6.5)}$$By substituting these 
expressions in Eq.~(6.4) and minimizing over the variable $r,$ we establish 
the upper bound (1.4) of Theorem~1. 
 
\title{VII.~~Summary and Conclusion} 
 
This paper is devoted to the investigation of the ground-state eigenvalue 
of~the semirelativistic (``spinless-Salpeter'') Hamiltonian (1.1) which 
governs the dynamics of a system of $N$ identical bosons that experience pair 
interactions described by a harmonic-oscillator potential with coupling 
strength $\gamma$. For a fixed coupling $\gamma=1,$ we have represented the 
exact ground-state energy eigenvalue of the corresponding \hi{1}{body} 
problem, regarded as a function $e(m)$ of the boson mass $m,$ by a monotone 
rising function $P(m),$ which is bounded by $1.376<P(0)\le P(m)\le 
P(\infty)=1.5.$ Our bounds (1.4) on the energy of the \hi{N}{body} problem 
are expressed in terms of~a formula which has this function $P$ as a 
parameter. 
 
In Fig.~2 we have plotted the energy bounds corresponding to fixed lower and 
upper limiting values of $P(m),$ namely, $P=\{1.376, 1.5\}.$ In Fig.~3 we 
have kept~the same upper energy bound, obtained with the help of a Gaussian 
trial wave function and corresponding to $P=1.5,$ but added the best lower 
energy bound of this type, using a ``running'' $P=P(\mu),$
 $\mu = m(N/(\gamma(N-1)^2))^{1\over 3}.$ The lower energy 
bound of Fig.~3 is identical to the exact energy for the case $N=2.$ For 
higher $N>2,$ Fig.~3 shows the approach of both upper and lower bounds to the 
well-known exact nonrelativistic solution (1.3) in the \hy{large}{m} limit. 
 
A key ingredient in this analysis is the use of relative coordinates: only in 
such a framework could the upper and lower energy bounds be made to converge 
in the Schr\"odinger limit. This study of the semirelativistic 
harmonic-oscillator problem is~a first step towards energy bounds valid for 
more general central pair interactions.\medskip 
 
\title{Acknowledgements} 
 
We thank W.~Thirring for a discussion about the \hi{N}{body} problem and 
G.~Dafni for confirming the reasoning in the proof of Lemma~1. Partial 
financial support of this work under Grant No. GP3438 from the Natural 
Sciences and Engineering Research Council of Canada, and hospitality of the 
Erwin Schr\"odinger International Institute for Mathematical Physics in 
Vienna is gratefully acknowledged by one of~us [R.~L.~H.].\medskip 
 
\np 
\references{1} 
\baselineskip 18 true pt 
 
\np\hbox{\vbox{\psfig{figure=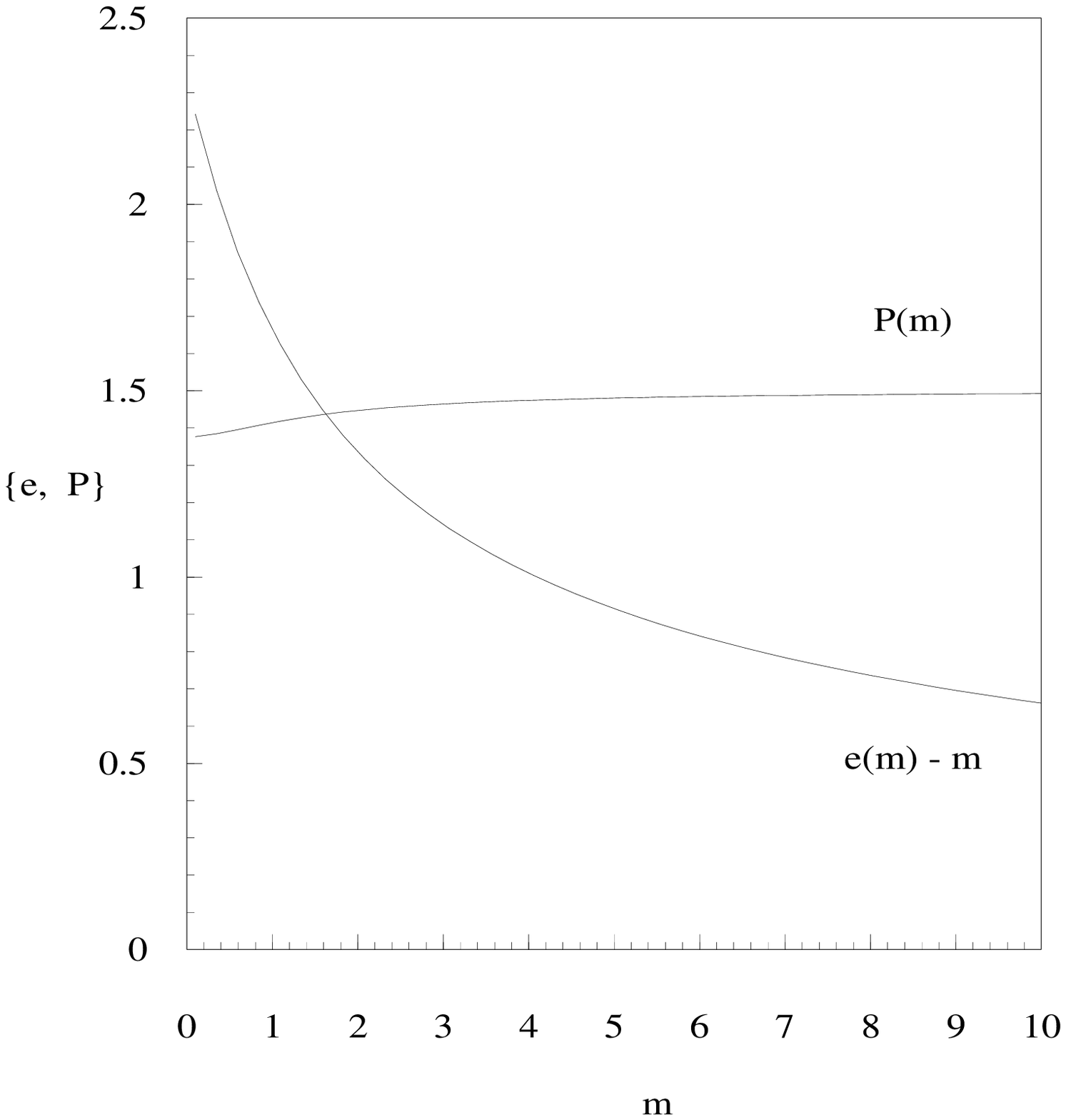,height=6in,width=5in,silent=}}}
\nl{\bf Figure 1.}~~The monotone energy function $e(m)$ of the \hi{1}{body} problem 
defined~by (3.1), and the monotone function $P(m)$ used in our standard 
representation (3.8) for $e(m);$ the function $P(m)$ is bounded by 
$P(0)=1.376\le P(m)\le P(\infty)=3/2.$ 
 
\np\hbox{\vbox{\psfig{figure=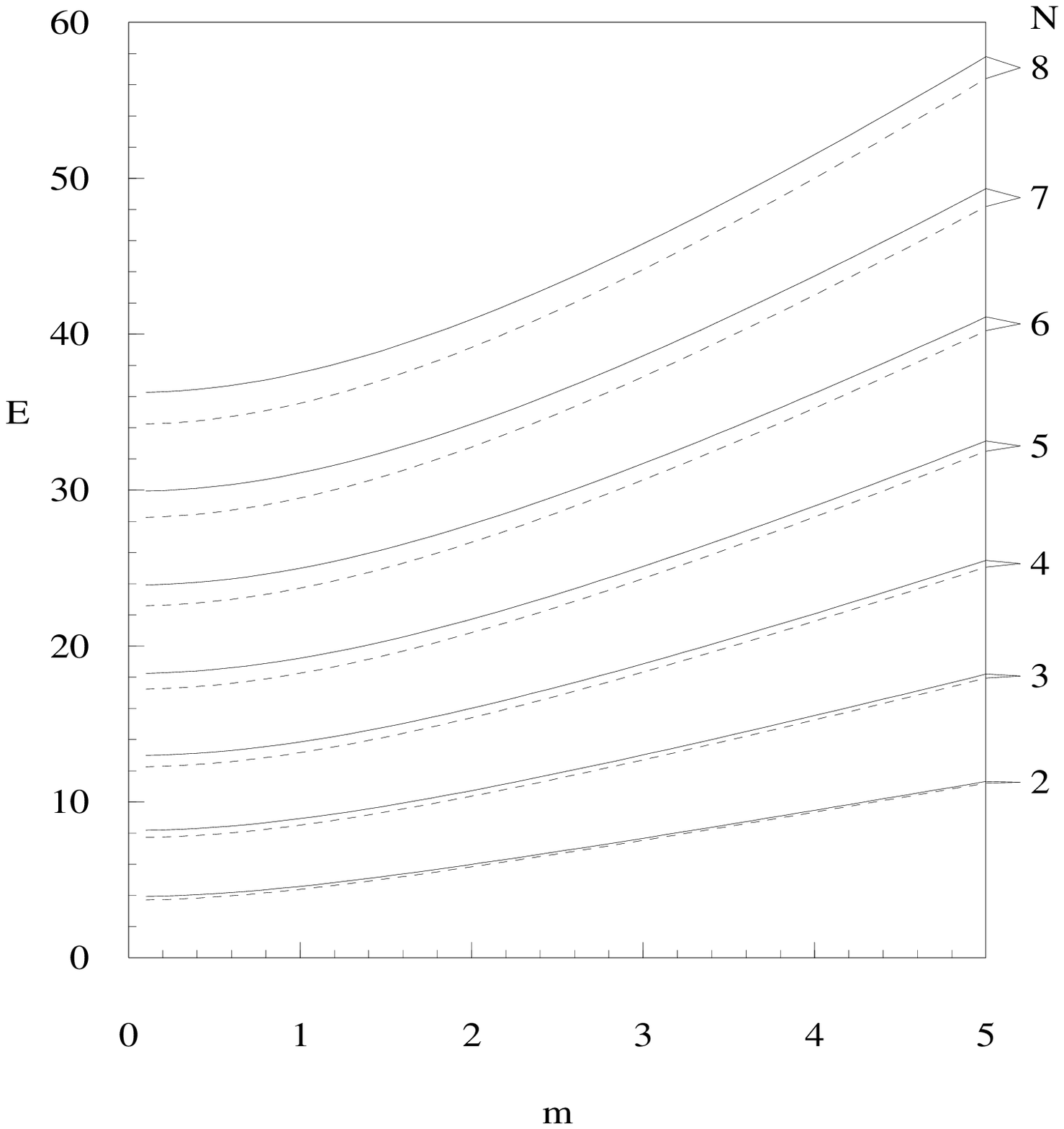,height=6in,width=5in,silent=}}}
\nl{\bf Figure 2.}~~Upper (full lines) and lower (dashed lines) bounds to the lowest 
energy $E(m)$ of the \hi{N}{boson} relativistic harmonic-oscillator problem 
for $N=2,3,\dots,8$ obtained by employing the constant values $P = 1.376$ and $P=1.5,$ 
respectively, in Eq.~(1.4) of Theorem~1.  
\np\hbox{\vbox{\psfig{figure=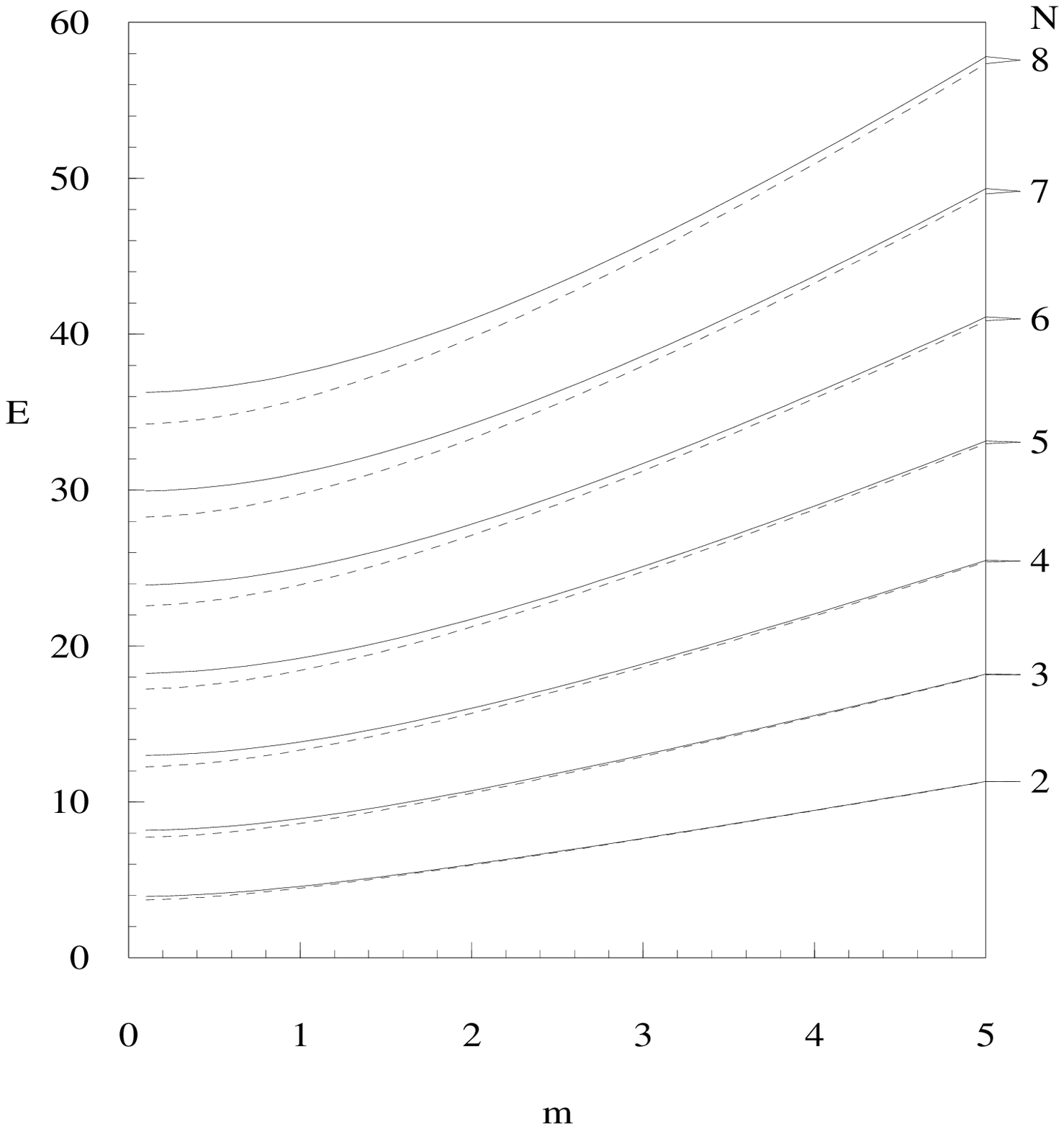,height=6in,width=5in,silent=}}}
\nl{\bf Figure~3.}~~Upper (full lines) and lower (dashed lines) bounds to the lowest 
energy $E(m)$ of the \hi{N}{boson} relativistic harmonic-oscillator problem 
for $N=2,3,\dots,8$ obtained by employing the values $P=P(\mu),\quad \mu = m(N/(\gamma(N-1)^2))^{1 \over 3},$ and $P=1.5,$ 
respectively, in Eq.~(1.4) of Theorem~1. For $N = 2,$ the lower bound is exact. 
 
 
\end